\documentstyle[editedbook,epsfig,psfig,epsf]{mq}

\def\cm2{cm$^2$ }
\def\se1{s$^{-1}$ }

\newcommand{\gtsim}{\protect\raisebox{-0.5ex}{$\:\stackrel{\textstyle >}{\sim}\:$}}

\begin{opening}
\title{TeV Neutrinos from Galactic Microquasar Jets}
\author{D. Guetta$^1$, C. Distefano$^{2}$,  A. Levinson$^3$ \& E. Waxman$^4$}
\institute{$^1$ Osservatorio di Arcetri, L.E. Fermi 2, Firenze\\
$^2$ Laboratori nazionali del Sud, INFN, Catania\\
$^3$ School of Physics and Astronomy, Tel Aviv University, Tel Aviv 69978, 
Israel\\
$^4$  Department of Condensed Matter Physics, Weizmann
Institute, Rehovot 76100, Israel.}
\end{opening}

\runningtitle{TeV neutrinos from Galactic microquasar jets, flux predictions.}
\runningauthor{Guetta et al.}

\begin{document}
\vspace{-0.5cm}
\begin{abstract}
{\small We discuss the possibility that microquasar jets may be powerful 
emitters of TeV neutrinos.  We estimate the neutrino fluxes 
produced by photopion production in the jets of
a sample of identified microquasars and microquasar candidates, for
which available data enables rough determination of the jet
parameters. We demonstrate that in several of
the sources considered, the neutrino flux at Earth, produced in
events similar to those observed, can exceed the detection
threshold of a km$^2$ neutrino detector. 
Sources with bulk Lorentz factors larger than
those characteristic of the sample considered here, directed along
our line of sight may be very difficult to
resolve at radio wavelengths and hence may be difficult to
identify as microqusar candidates.
However these sources can  be identified through their
neutrino and gamma-ray emission.}
\end{abstract}

\section{Introduction}
The composition of microquasars jets is yet an open issue.
The synchrotron emission both in the radio and in the IR is consistent
with near equipartition between electrons and magnetic field,
which is also implied by minimum energy considerations
\cite{Levinson96}. However, the dominant energy carrier in the
jet is presently unknown (with the exception of the jet in SS433).
A possible diagnostic of hadronic jets is emission of TeV neutrinos 
\cite{Levinson01}.  As shown in reference \cite{Levinson01}, for typical microquasar
jet parameters, protons may be accelerated in the jet to energies in excess
of $\sim10^{16}$~eV.  The interaction of these protons with
synchrotron photons emitted by thermal electrons is
expected to lead to 1--100~TeV neutrino emission. The predicted
fluxes are detectable by large, km$^2$-scale effective area,
high-energy neutrino telescopes, such as the operating south pole
detector AMANDA \cite{Andres00} and its planned 1~km$^2$
extension IceCube \cite{icecube}, or the Mediterranean sea
detectors under construction (ANTARES \cite{antares};  NESTOR \cite{nestor}) 
and planning (NEMO  \cite{nemo} and \cite{Halzen01} for a recent review).

In this paper we consider a class of identified Galactic
microquasars with either persistent jets or documented outbursts.
For each source we provide, for illustrative purposes, our model
prediction for the neutrino flux that should have been emitted
during particular events, using radio data available in the
literature. Although the temporal behavior of many of these
sources may be unpredictable, we demonstrate that some of the
sources could have been detected by a neutrino telescope with
effective area larger than km$^2$ (in some cases even 0.1 km$^2$)
had such a detector been in operation during the time of the
recorded events and, therefore, propose that they should be
potential targets for the planned neutrino telescopes.  In
addition we consider a list of XRBs thus far unresolved at radio
wavelengths, that are believed to be microquasar candidates.
In \S\ref{sec:model} we briefly discuss the neutrino production
mechanism in microquasars. In \S\ref{sec:L_jet} we use
observational data available for each source to estimate the jet
parameters, and then use these parameters to derive the expected
neutrino flux. The number of neutrino induced muon events in
km$^2$-scale neutrino telescopes is derived in \S\ref{sec:N_mu}.
The implications of our results are briefly discussed in
\S\ref{sec:discussion}.

\section{Internal shock model for microquasars}
\label{sec:model}

In this section we give a brief outline of the model
proposed by Levinson \& Waxman (2001) \cite{Levinson01}
for production of neutrinos in microquasars
with the purpose to introduce the parameters relevant for the present
analysis. 
The model assumes that on sufficiently small
scales ($\leq 10^{11}$ cm), unsteadiness of the jet leads to the formation
of internal shocks that can accelerate protons and electrons to a
power law distribution. For typical jet parameters,
the maximum proton energy is roughly $10^{16}$~eV in the jet frame. 
Protons can interact with both the external X-ray photons emitted by the
accretion disc, and the synchrotron photons produced inside
the jet by the shock heated electrons, leading to pion production
and consequent neutrino emission. In order for photomeson
production to take place, the comoving proton energy must exceed
the threshold energy for $\Delta$-resonance, $\approx10^{14}$ eV
for interaction with the external photons and $\approx10^{13}$ eV
for interaction with synchrotron photons.

Charged pions produced in photo-meson interactions decay to
produce neutrinos, $\pi^+ \rightarrow \mu^+\ +\ \nu_{\mu}
\rightarrow e^+ \ +\ \nu_e\ +\ \bar{\nu}_\mu+\ \nu_{\mu}$. In a
single collision, a pion is created with an average energy that
is $\approx20\%$ of the proton energy. This energy is roughly
evenly distributed between the final $\pi^+$ decay products,
yielding a $\nu$ energy that is $\approx5\%$ of the proton
energy. The neutrino signal is dominated by
neutrinos in the energy range of 1 to 100~TeV.
The fraction of proton energy converted into pions, $f_{\pi}$,
depends on the jet Lorentz factor, $\Gamma$, and on  the kinetic
luminosity of the jet, $L_{jet}$ (see \cite{distefano} for details).

The expected neutrino flux (energy per unit area per unit time)
of $\gtsim1$~TeV muon neutrinos at Earth from a jet ejection event is
\cite{Levinson01}
\begin{equation}
f_{\nu_\mu}\simeq \frac{1}{2}\eta_p
f_\pi\delta^4\frac{L_{jet}/8}{4\pi D^2}.
\label{eq:flux_nu}
\end{equation}
where $\delta=[\Gamma(1-\beta\cos\theta)]^{-1}$ is the Doppler
factor of the jet ($\theta$ is the angle between the jet axis and
our line of sight), $\eta_p\sim0.1$ is the fraction of $L_{jet}$
carried by accelerated protons, and $D$ is the source distance.

\section{Jet parameters and neutrino fluxes}
\label{sec:L_jet}

In order to determine the neutrino flux from (\ref{eq:flux_nu})
the main quantity to be estimated is the kinetic power of the jet.
We have used two different methods for
the sources with the relativistic jets resolved in the radio band
(resolved microquasars) and for the XRBs observed as point like sources,
such as GX 339-4, henceforth referred to as unresolved microquasars \cite{Fender01}. 

\subsection{Resolved microquasars}

In events that have been monitored with sufficiently good resolution,
it is often possible to obtain a rough estimate of the
characteristic source parameters, in particular the bulk speed of
the jet, the angle between the jet axis and the sight line, and
the size of the emitting blob.
$L_{jet}$ is
estimated in the following way: for a flux density
$S_\nu\propto\nu^{-1/2}$ of the radio source, implying an electron
energy distribution $dn_e/d\epsilon_e\propto\epsilon^{-2}$, we estimate the
minimum energy carried by electrons and magnetic field, obtained
for a magnetic field [e.g. \cite{Levinson96}]:
\begin{equation}
B_{*}=3.6\left[\ln(\gamma_{\rm max}/\gamma_{\rm min})
\frac{T_{B6}}{l_{15}} \right]^{2/7}
\frac{\nu^{5/7}_9}{\delta}\quad {\rm mG},
\label{eq:B_ep}
\end{equation}
where $\nu=\nu_9\cdot10^9$~Hz, $l=l_{15}\cdot10^{15}$~cm is the size of
the emission region, $\gamma_{\rm max}$ and $\gamma_{\rm min}$ are
the maximum and minimum electron random Lorentz factors as
measured in the jet frame, and $T_B=10^6T_{B6}$ K is the
brightness temperature.
The jet power then satisfy,
\begin{equation}
L_{jet}\ge 0.3 c (\Gamma l B_{*})^2.
\label{eq:L_jet}
\end{equation}
For the numerical estimates that follow we conservatively assume
$\gamma_{\rm max}/\gamma_{\rm min}=100$.
In Table \ref{tab:tab} we report our estimates of the jet power and of 
the neutrino flux expected at Earth, calculated using equations 
\ref{eq:flux_nu} and \ref{eq:L_jet}
and adopting the source parameters quoted in literature.
The two values of the periodic source LS I
+61$^\circ$303 (P$\sim$26.5 days) refer respectively to bursting
and quiescent states observed by Massi et al. (2001) \cite{Massi01}.  
The distance of V4641 Sgr, and therefore the jet parameters, is uncertain.
We report calculations for both the values of $D$ present in literature 
\cite{Hjellming00} and \cite{Orosz01}. 
Several authors pointed out that the kinetic energy output of the
SS433 jets can influence the radio emission of W50, moreover SS433
is the only microquasar that shows a strong H$_\alpha$-line
emission from the jets \cite{Begelman80,Davidson80,Kirshner80}.
In our estimate we assume the conservative value of $L_{jet}\sim
10^{39}$ erg/sec suggested by Margon (1984)\cite{Margon84}.

\subsection{Unresolved microquasars}

For the sources whose jet has not been resolved, we cannot deduce
the value of $L_{jet}$ from Eq.(\ref{eq:L_jet}), since $\Gamma$,
$\delta$ and the size of the jet are not known. We follow instead
a different line of argument. 
We estimate the jet synchrotron luminosity, assuming a spectral index
$\alpha_R\sim 0.5$ for the emitted radiation \cite{Fender01}:
\begin{equation}
L_{syn}= 4\pi D^2 \frac{1}{1-\alpha_R} S_{\nu_{high}} \nu_{high},
\label{eq:L_rad}
\end{equation}
where  $\nu_{high}$ is the highest observed frequency of synchrotron emission, and  
$S_{\nu_{high}}$ the flux density emitted at this frequency.
Assuming that the total jet synchrotron luminosity is a fraction $\eta_r$ of the
jet kinetic energy, carried by relativistic electrons and magnetic field, we
estimate the total Ljet as: 
\begin{equation}
L_{jet}= \eta_e^{-1}\eta_r^{-1} L_{syn},
\label{eq:Ljet2}
\end{equation}
where $\eta_e$ denotes the fraction of bulk energy converted internal energy of electrons.  
For a typical value of $\alpha_R\sim 0.5$, that implies electrons do not cool fast on
scales that are resolved by the VLA. Assuming that emission at $\nu_{high}$ originates
from the same radius as the radio emission, for the typical parameters inferred
for the resolved microquasars (B of tens of mG corresponding synchrotron frequency of
about 10 GHz), electrons radiating  at  $\nu_{high}\sim10^{14}$ cannot lose more 
than $\eta_r\sim0.1$ of their energy. The energy neutrino flux expected at Earth is:
\begin{equation}
f_{\nu_\mu} =  \frac{1}{2}f_\pi \eta_p \frac{L_{jet}/8}{4\pi D^2}=
\frac{1}{16(1-\alpha_R)}\frac{\eta_p}{\eta_e} f_\pi \eta_r^{-1}
S_{\nu_{high}} \nu_{high}.
\label{eq:F_nu2}
\end{equation}
We have neglected in eqs. (\ref{eq:L_rad}--\ref{eq:F_nu2})
corrections associated with the relativistic expansion of the jets. However,
since these corrections are the same for both the synchrotron and
neutrino emission, the estimate of neutrino flux in eqs.
(\ref{eq:F_nu2}) is independent of such corrections.
In Table \ref{tab:tab} we quote, for a sample of unresolved
microquasar candidates, our estimates of
$L_{jet}$ and $f_{\nu_\mu}$, calculated from Eq. (\ref{eq:Ljet2})
and Eq.(\ref{eq:F_nu2}).

Recent observations with the VLBA array have confirmed the nature of microquasars
for the source Cygnus X-1, revealing an extended jet-like feature extending to $\sim$
15 mas from a core region, with an opening angle of $<2^\circ$ \cite{Stirling01}. 
The authors suggest for this source a bulk motion velocity of $\beta\sim0.75$ and a
viewing angle of $\theta\sim 40^\circ$, corresponding to a Lorentz factor of
$\Gamma\sim1.51$ and a Doppler boost factor $\delta\sim 1.55$.
Adopting these jet parameters and a quasi-steady emission of $S_\nu\sim5$ mJy at
$\nu=8.4$ GHz, the kinetic luminosity,
calculated from Eq. (\ref{eq:L_jet}), is $L_{jet}\sim3.53\cdot10^{36}$ erg/sec.
The new estimate of the $L_{jet}$ for Cygnus X-1 is therefore consistent with the
value in Tab. \ref{tab:tab}, obtained from the Eq. (\ref{eq:Ljet2}) for the
unresolved sources.

\section{Muon events expected in a km$^2$-scale detector}
\label{sec:N_mu}

The detection of TeV neutrino fluxes from microquasars could be
the first achievable goal for proposed underwater(ice) neutrino
telescopes. In this section we calculate the rate of
neutrino-induced muon events expected in a detector with an
effective area of $1~{\rm km}^2$. Since the
signal is expected to be dominated by neutrinos of energy
$E_\nu\ge1$~TeV, for which the detection probability is
$P_{\nu\mu}\sim 1.3 \cdot 10^{-6} E_{\nu,TeV}$ \cite{Gaisser95},
we estimate the number of neutrino-induced muon events as
\cite{Levinson01}:
\begin{equation}
N_\mu \simeq 0.2 \eta_{p,-1}f_\pi\delta^4 D_{22}^{-2}
L_{jet,38} A_{{\rm km}^2}\Delta t_d
\label{eq:Mu_rate}
\end{equation}
where $A_{{\rm km}^2}$ is the effective detector area in km$^2$
units, and $\Delta t_d$ the duration of the observed burst measured in days.
{\em It is seen from eq. \ref{eq:Mu_rate} that in microblazars (microquasars
having their jets pointing in our direction), even single outbursts may be
easily detectable by km$^2$ detectors if the total energy release
is of the order of that seen in GRS 1915 and GRO 1655
($L_{jet}\Delta t\sim 10^{43}$ ergs). }
In Table
\ref{tab:tab} we report the number of events expected in a
detector with $A_{eff}=1~{\rm km}^2$, during the bursts considered
in section \ref{sec:L_jet}. In the case of persistent sources
number of neutrino induced muon events 
 is calculated for a 1 year period. In the same
table, we also report the number of background atmospheric neutrino events
collected in such a detector during the time $\Delta t$,
assuming a neutrino spectrum
$\phi_{\nu,bkg} \sim  10^{-7} E^{-2.5}_{\nu,\rm TeV}/ {\rm
cm}^2~{\rm sec}~{\rm sr}$ for $E_\nu>1$~TeV and
for a detector angular resolution of  0.3$^\circ$.

\section{Discussion}
\label{sec:discussion}

There are large uncertainties involved in the derivation of the
jet parameters for most of the sources listed in table 4. The best
studied cases are perhaps GRS 1915+105, GRO J1655-40, and SS433.
Nonetheless, we have demonstrated that if the jets in microquasars
are protonic, and if a fraction of a few percent of the jet energy
is dissipated on sufficiently small scales, then emission of TeV
neutrinos with fluxes in excess of detection limit of the
forthcoming, km$^2$ scale, neutrino telescopes is anticipated.

The present identification of microquasars, and the inferred
distribution of their jet Lorentz factors, may be strongly
influenced by selection effects. It is quite likely that the class
of Galactic microquasars contains also sources with larger bulk
Lorentz factors and smaller viewing angles, which should emit
neutrinos with fluxes considerably larger than the extended
microquasars discussed in this paper.  Such sources may be
identified via their gamma-ray emission with, e.g. AGILE and GLAST, or by
their neutrino emission. The gamma-rays should originate from
larger scales where the pair production opacity is sufficiently
reduced. Predictions for AGILE and GLAST will be discussed 
in a forthcoming paper \cite{Guetta02}. There
are currently about 280 known XRBs \cite{Liu00,Liu01}, of
which $\sim 50$ are radio loud. These may also be potential
targets for the planned neutrino detectors.
Our results quoted in Table \ref{tab:tab} 
are consistent with experimental upper limits on neutrino fluxes
from point sources set by MACRO \cite{Ambrosio01}, AMANDA
\cite{Biron} and SuperKamiokande \cite{Okada00}.

\begin{table*}
\caption{Estimates of the kinetic luminosity of the jet and neutrino flux for known
microquasars. N$_\mu$ is the predicted number of muon events in a
$1{\rm km}^2$ detector during the burts duration $\Delta t$ and N$_{\mu,bkg}$ is the
expected number of background atmospheric neutrino events, assuming
angular resolution of 0.3$^\circ$. \label{tab:tab}} 
\begin{center}
\begin{small}
\begin{tabular}{lccccc}
\hline \hline
\multicolumn{6}{}{}\\
Source name & $L_{jet}$ & $f_{\nu_\mu}$ & $\Delta t$  & N$_\mu$& N$_{\mu,bkg}(deg/0.3^\circ)^2$      \\
	    & (erg/sec) & (erg/cm$^2$sec)& (days)     &         &    \\
\hline
Resolved \\
\hline
CI Cam 	\cite{Hjellming98,Harmon98}&      5.66$\cdot10^{36}$   & 2.23$\cdot10^{-11}$             & 0.6  & 0.005 & 0.002    \\
XTE J1748-288 \cite{Rupen98,Mirabel99,Kotani00}	&  1.84$\cdot10^{38}$   & 3.07$\cdot10^{-11}$           & 20   & 0.25 & 0.054     \\
Cygnus X-3 \cite{Mioduszewski01}  	&  1.17$\cdot10^{38}$   & 4.02$\cdot10^{-10}$            & 3    & 0.48 & 0.008     \\
LS 5039  \cite{Paredes00}    	&   8.73$\cdot10^{35}$   & 1.69$\cdot10^{-13}$           & persistent & 0.02 & 0.986  \\
GRO J1655-40 \cite{Hjellming95}	&  1.60$\cdot10^{39}$   & 7.37$\cdot10^{-11}$            & 6    & 0.18 & 0.016     \\
GRS 1915+105 \cite{Mirabel94}  	&  2.45$\cdot10^{39}$   & 2.10$\cdot10^{-11}$           & 6    & 0.05 & 0.016     \\
Circinus X-1 \cite{Mirabel99,Preston83} 	&   7.61$\cdot10^{37}$   & 1.22$\cdot10^{-11}$         & 4    & 0.02 & 0.011     \\
LS I 61$^\circ$303 \cite{Massi01}& 1.65$\cdot10^{36}$   & 4.49$\cdot10^{-12}$ 	& 7    & 0.01 & 0.019     \\
LS I 61$^\circ$303 \cite{Massi01}& 5.69$\cdot10^{35}$   & 9.06$\cdot10^{-13}$ 	& 20   & 0.01 & 0.054     \\
XTE J1550-564 \cite{Hannikainen01,Sanches99}	&   2.01$\cdot10^{37}$   & 2.00$\cdot10^{-12}$         & 5    & 0.004 & 0.014    \\
V4641 Sgr \cite{Hjellming00}   	&     8.02$\cdot10^{36}$   & 2.25$\cdot10^{-11}$         & 0.3  & 0.003 & 0.001    \\
V4641 Sgr \cite{Hjellming00,Orosz01}    	&  1.17$\cdot10^{39}$   & 3.25$\cdot10^{-9}$          & 0.3  & 0.39 & 0.001     \\
Scorpius X-1 \cite{Fomalont01a,Fomalont01b}	&  1.04$\cdot10^{37}$   & 6.48$\cdot10^{-13}$ 		& persistent & 0.09 & 0.986 \\
SS433 \cite{Margon84}      	&   10$^{39}$            & 1.72$\cdot10^{-9}$           & persistent & 252 & 0.986 \\
\hline
Unresolved \\
\hline
GS 1354-64 \cite{Fender00,Brocksopp01}	&  3.62$\cdot10^{37}$ & 1.88$\cdot10^{-11}$ 	& 2.8  & 0.02 & 0.008      \\
GX 339-4 \cite{Fender00,Fender01b,Corbel00}	&  3.86$\cdot10^{38}$ & 1.26$\cdot10^{-9}$ 	& persistent & 183.4 & 0.986 \\
Cygnus X-1\cite{Fender01b,Pooley99}	&  1.45$\cdot10^{36}$ & 1.88$\cdot10^{-11}$ 	& persistent & 2.8 & 0.986  \\
GRO J0422+32 \cite{Fender00,Shrader94}	&  4.35$\cdot10^{37}$ & 2.51$\cdot10^{-10}$ 	& 1$\div$20   & 0.1$\div$2 & 0.003$\div$0.1 \\
XTE J1118+480 \cite{Fender00,Frontera01}	&  3.49$\cdot10^{37}$ & 5.02$\cdot10^{-10}$ 	& 30$\div$150 & 6$\div$30 & 0.081$\div$0.405   \\
\hline \hline
\end{tabular}
\end{small}
\end{center}
\end{table*}

\end{document}